\documentclass[10pt,conference]{IEEEtran}
\IEEEoverridecommandlockouts
\usepackage{algorithm}
\usepackage{algorithmic}
\usepackage{amsfonts}
\usepackage{amsmath}
\usepackage{color}
\usepackage{esvect}
\usepackage{gensymb}
\usepackage{graphicx}
\usepackage{mathtools}
\usepackage[binary-units=true]{siunitx}
\usepackage{stfloats}
\usepackage{url}
\usepackage{xspace}
\usepackage{wrapfig}
\usepackage{braket}
\usepackage{siunitx}
\usepackage[normalem]{ulem}
\usepackage{enumitem}
\usepackage{subcaption}
\usepackage[sort,square,numbers]{natbib}

\makeatletter
\newcommand\fs@spaceruled{\def\@fs@cfont{\bfseries}\let\@fs@capt\floatc@ruled
  \def\@fs@pre{\vspace{0.5\baselineskip}\hrule height.8pt depth0pt \kern2pt}%
  \def\@fs@post{\kern2pt\hrule\relax\vspace{-0.4cm}}%
  \def\@fs@mid{\kern2pt\hrule\kern2pt}%
  \let\@fs@iftopcapt\iftrue}
\makeatother
\floatstyle{spaceruled}
\restylefloat{algorithm}

\newcommand{\eat}[1]{}

\newcommand{\blue}[1]{{\color{blue}#1}}

\newcommand{\para}[1]{\smallskip \noindent {\bf #1}}
\newcommand{\softpara}[1]{\smallskip \noindent \underline{#1}}

\newcommand{\showOutline}{s}
\newcommand{\hide}[1]{\blue{h}}

\definecolor{darkgreen}{rgb}{0,0.5,0}
\definecolor{darkblue}{rgb}{0,0,0.5}

\newcommand{\outline}[2]{\if#1\showOutline {\color{darkgreen}#2} \fi}
\newcommand{\detailedoutline}[2]{\if#1\showOutline {\color{darkblue}#2} \fi}

\newcommand{\todo}[2]{\if#1\showOutline {#2} \fi}

\newcommand{\purp}[1]{} 

\def\blackbox{\hfill {\vrule height6pt width6pt depth0pt}}
\newcommand{\vsplem}{\vspace{0.025in}}
\newcommand{\vsp}{\vspace*{-0.1in}} 

\newtheorem{lem}{Lemma}
\newtheorem{thm}{Theorem}
\newtheorem{obv}{Observation}

\newenvironment{lem-wo-prf}{\begin{lem} \nopagebreak}{{\hfill$\blackbox$} \end{lem}}
\newenvironment{lem-w-prf}{\vsp \begin{lem} \nopagebreak}{\end{lem}}
\newenvironment{thm-w-prf}{\vsp \begin{thm} \nopagebreak}{\end{thm}}
\newenvironment{thm-wo-prf}{\begin{thm} \nopagebreak}{{\hfill$\blackbox$} \end{thm}}

\newcounter{packednmbr}

\newcommand{\eps}{\mbox{EP}\xspace}
\newcommand{\epss}{\mbox{EPs}\xspace}

\newcommand{\spsw}{{\tt SP-SWO}\xspace}
\newcommand{\mpsw}{{\tt MP-SWO}\xspace}
\newcommand{\EP}{{\tt EP}\xspace}

\newcommand{\ES}{{\tt ES}\xspace}
\newcommand{\ESs}{{\tt ESs}\xspace}

\newcommand{\EPs}{{\tt EP}s\xspace}
\newcommand{\DP}{\textsc{DP}\xspace}
\newcommand{\genDP}{\textsc{DP-with-ages}\xspace}
\newcommand{\midDP}{\textsc{swap-or-wait}\xspace}
\newcommand{\simDP}{\textsc{DP}\xspace}
\newcommand{\MDP}{\textsc{MDP}\xspace}
\newcommand{\Asap}{\textsc{Swap-ASAP}\xspace}
\newcommand{\LS}{\textsc{Greedy-SP}\xspace}
\newcommand{\LP}{\textsc{LP}\xspace}
\newcommand{\LSMP}{\textsc{Greedy-MP}\xspace}

\newtheorem{defin}{Definition}

\newcommand{\php}{\mbox{$p_{ob}$}\xspace}          
\newcommand{\bt}{\mbox{$t_{b}$}\xspace}         
\newcommand{\bp}{\mbox{$p_{b}$}\xspace}          
\newcommand{\gt}{\mbox{$t_g$}\xspace}      
\newcommand{\gp}{\mbox{$p_g$}\xspace}       
\newcommand{\ep}{\mbox{$p_e$}\xspace}       
\usepackage[font=small]{caption}
\title{Optimized Generation of Entanglement by Real-Time Ordering of Swapping Operations\vspace{-0.1in}}

\author{\IEEEauthorblockN{Ranjani G. Sundaram}
\IEEEauthorblockA{
\textit{ Stony Brook University, NY}}
\and
\IEEEauthorblockN{Himanshu {Gupta}}
\IEEEauthorblockA{
\textit{ Stony Brook University, NY}}}
\IEEEaftertitletext{\vspace{-3\baselineskip}}

\begin{document}
\maketitle
\pagestyle{plain}
\begin{abstract}
Long-distance quantum communication in quantum networks faces significant challenges due to the constraints imposed by the no-cloning theorem. Most existing quantum communication protocols rely on the a priori distribution of entanglement pairs (\EPs), a process known to incur considerable latency due to its stochastic nature. In this work, we consider the problem of minimizing the latency of establishing an EP across a pair of nodes in a quantum network.
While prior research has primarily focused on minimizing the expected generation latency by selecting {\em static} entanglement routes and/or swapping trees in advance, our approach considers a real-time adaptive strategy---wherein the order of entanglement-swapping operations (hence, the swapping tree used) is progressively determined at runtime based on the runtime success/failure of the  stochastic events. In this context, we present a greedy algorithm that iteratively determines the best route and/or entanglement-swapping operation to perform at each stage based on the current network.
We evaluate our schemes on randomly generated networks and observe a reduction in latency of up to 40\% from the optimal offline approach.
\end{abstract}

\section{\bf Introduction}

Quantum networks enable the construction of large, robust, and more capable
quantum computing platforms by connecting smaller QCs. 
Quantum networks \cite{simon2017towards} also enable various important applications \cite{eldredge2018optimal, komar2014quantum}. 
Quantum network communication is challenging, as the physical transmission of quantum states across nodes can incur irreparable communication errors
due to quantum no-cloning \cite{dieks1982communication}. 
Thus, quantum protocols such as teleportation~\cite{bennett1993teleporting} or telegates~\cite{Telegates} are used to facilitate quantum communication or computation across nodes in a quantum network; these protocols, however, require an a priori entangled pair (\EPs) shared across the end nodes. 
Our work focuses on efficiently generating \EPs in a quantum network.

Generating \EPs over long distances in a quantum network is challenging. 
In particular, coordinated entanglement swapping (e.g., DLCZ protocol \cite{duan2001long}) using quantum repeaters can be used to establish
long-distance entanglements from short-distance entanglements. 
However, due to the low probability of success of the underlying physical processes, \EP generation can incur significant latency~\cite{sangouard2011quantum}; high generation latency of EPs can 
lead to high circuit execution times and even defeat their purpose due to decoherence.
Thus, this paper considers the problem of generating \EPs across remote nodes in a quantum network with minimal latency. 


\para{Adaptive Generation of EPs.}
Prior work on efficient \EP generation~\cite{Chakraborty2020, ShiQian, Pant17,Chakraborty2019,DPpaper, ghaderibaneh2023generation}  almost exclusively consists of {\em offline} approaches that select {\em static} entanglement routes or swapping trees to minimize \emph{expected} generation latency. 
These structures are constructed based on expected latencies and, thus, may incur high latency at runtime depending on the actual outcomes of the underlying stochastic events, as illustrated later (Example~\ref{fig:StaticvDynamic}). 
In contrast, a runtime-adaptive approach that selects the entanglement route and/or the order of entanglement-swapping operations 
based on the outcomes of the stochastic events {\em during the generation process} can potentially minimize the runtime generation latency.
Moreover, such adaptive approaches can also be very effective in incorporating decoherence constraints due to the runtime availability of 
actual ages of the qubits. 
Based on the above insights, in this paper, we develop effective adaptive generation approaches to generate EPs with minimal latency in quantum networks. 
To the best of our knowledge, the only prior work that develops an adaptive 
EP generation approach is~\cite{MDP}---but they assume a {\em given} entanglement path, and the time and space complexity of their approach is exponential 
in the network size and thus feasibly only for very small (at most 5-10 nodes) networks and low decoherence times (see \S\ref{sec:eval}).

\para{Our Contributions.} In the above context, we make the following contributions.
\begin{itemize}
    \item We first consider a simplified version of the problem where entanglement-swapping operations can be performed along exactly one selected (entanglement) path in the network. In this setting, we design an adaptive algorithm that, at each stage, selects the best entanglement-swapping operation to be performed based on the current network state (available EPs and their ages).
    
    \item We also consider a generalized setting wherein the designed adaptive approach selects, at each stage, the best route and the entanglement-swapping operation to be performed based on the current network state. 
    \item We conduct extensive evaluations of our algorithms on a wide array of quantum networks and observe a reduction in latency of up to $40\%$ compared to prior static approaches. 
\end{itemize}

\section{\bf Background}
\label{sec:Background}

This section presents the relevant background related to the generation of \EPs over remote nodes in a quantum network.

\para{Quantum  Network (QN).}
We consider a quantum network (QN)
as a graph over the set of quantum computers as network nodes.
Each edge in the graph represents a (quantum and classical) communication 
link. We refer to these network edges as {\em network links} or just {\em links}.
Our network model is similar to the one used in 
some of the recent works~\cite{DPpaper,ghaderibaneh2023generation} on
efficient generation of \epss and graph states.
In particular, each node has an atom-photon \eps generator with 
latency (\gt) and probability of success (\gp).
A node's atom-photon generation capacity/rate 
is its aggregate capacity and may be split across its incident links.
Each network link generates \EPs (called {\em link-\EPs}) 
shared over the adjacent nodes 
using the nodes' atom-photon generators 
and an optical-BSM device located in the middle of the link. 
The optical-BSM  has 
a certain probability of success (\php), 
and each half-link (from either end-node)
to the device has a probability of transmission success (\ep).
We implicitly assume the 
atom-photon generation latency to include
the times for photon transmission, 
optical-BSM latency, and classical acknowledgment; thus, the 
the generation latency of a link-\EP is $\gt/(\gp^2\ep^2\php)$ 
between two nodes.
To facilitate atom-atom \ES operations, 
each network node is also equipped 
with an atomic-BSM device 
with appropriate operation latency \bt and
probability of success \bp.

\para{\bf Generating \EPs over Remote Nodes.}
To distribute an \EP over remote nodes, \EPs are
generated over larger and larger distances from shorter-distance \EPs using a sequence of ``entanglement-swapping'' operations. 
An entanglement swapping (\ES) operation can be looked up as being performed over three nodes $(A, B, C)$ with two \EPs over $(A, B)$ and $(B,C)$; the 
\ES operation results in an \EP over the nodes $(A, C)$, by essentially teleporting the first qubit in node $B$ (i.e., the qubit of the EP over $(A,B$)) to the node $C$ using the second \EP over $(B,C)$.
In general, an \EP over a pair of remote nodes $A$ and $B$ can be generated by
a sequence of \ES operations over the \EPs over adjacent 
nodes along a path from $A$ to $B$.
\begin{figure}
\centering
    \includegraphics[width=0.4\textwidth]{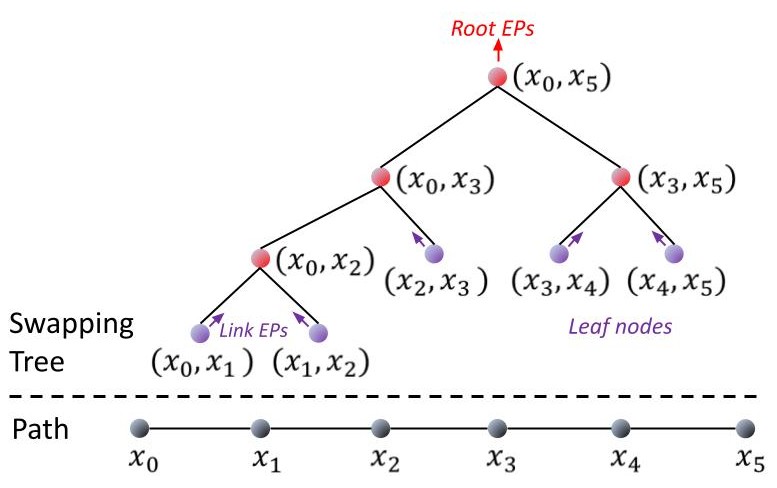}
    \caption{Establish \EPs over remote nodes.}
    \label{fig:swapping_tree}
  \vspace{-0.5cm}
\end{figure}

\softpara{Swapping Trees.} In recent work, \cite{DPpaper} introduces the concept of \emph{swapping trees}, which are a characterization of the different orderings of \ES operations that can be used to establish an \EP over the endpoints of a network path.  Given a path $P$ with endpoints $s$ and $d$, any complete binary tree over the ordered links in $P$ gives a way to generate an \EP over $(s,d)$. Each vertex in the tree corresponds to a pair of network nodes in $P$, with each leaf representing a link (See Fig~\ref{fig:swapping_tree}). The latency of a swapping tree is the \emph{expected} time taken to establish the \EP corresponding to the tree's root node. 
 
\para{Decoherence Threshold Time $\tau$.} 
The fidelity of the quantum state decreases over time due to interaction with the environment; this is called decoherence. In order to limit the decoherence of qubits in an EP, we enforce an upper bound on the age (time spent in a quantum memory) of the qubits involved. 
A qubit is said to satisfy the decoherence constraint if its age is below 
a decoherence threshold of $\tau$.


\section{\bf Problem Formulation and Related Work}
\label{sec:ProblemFormulation}

We start by formulating the real-time\footnote{We use real-time and runtime interchangeably, in this paper.} entanglement generation problem addressed in this paper. We now consider entanglement generation over a single path in the network; we will consider the more general multi-path problem later in~\S\ref{sec:MultiPath}.

\para{\spsw Problem (Single-Path Swap Ordering).} 
Given a quantum network and a source-destination pair $\{(s, d)\}$, 
the \spsw problem is to determine a path $P$ from $s$ to $d$ and a real-time (during \EP generation) ordering of the entanglement-swapping (\ES) operations  so as to minimize the actual generation latency of the {\em target} \EP $(s,d)$ under the below node and decoherence constraints.

\begin{enumerate}
    \item \textit{Node Constraints.} For each node, the aggregate resources used (in generating the link-EPs) are less than the available resources.
    \item \textit{Decoherence Constraints.} The age of any qubit is less than the decoherence threshold $\tau$.
\end{enumerate}

\para{Related Work}
In recent years, a significant amount of work has been done on  
efficiently generating remote \EPs in a quantum network. Below, we discuss
these works categorized as follows.

\softpara{Selection of Entanglement Routes.} 
Most of the work on efficient generation 
of \EPs in quantum networks focuses on selecting one or more entanglement paths/routes~\cite{Chakraborty2020, ShiQian, Pant17,Chakraborty2019}. These works
typically assume that the underlying processes, including link-\EP generation and atomic BSMs, are synchronized; if all of these processes along the chosen entanglement route succeed, the end-to-end \EP is generated.
On failure of any of the 
underlying processes (atomic BSM or link-\EP generation), all the generated link/subpath-\EPs are discarded, and the whole process starts again from scratch. 
Such approaches require no quantum memories or require only memories with minimal storage time.
In particular,~\cite{ShiQian} designs a Dijkstra-like algorithm 
to construct an optimal path between a pair of nodes when there are multiple 
channels between adjacent nodes; they generalize their techniques to also
select multiple paths over multiple pairs of nodes. In 
addition,~\cite{Chakraborty2020} designs a linear programming (LP)
formulation based on multi-commodity flow to select routing paths 
for a given set of source-destination pairs; they map the
operation-based fidelity constraint to the path length and use copies of
nodes to model the fidelity constraint in the LP.
Among earlier works,~\cite{Pant17} proposes a greedy solution
for grid networks, and~\cite{virtual} proposes 
virtual-path-based routing in ring/grid networks.


\softpara{Constructing Swapping Trees.}
In networks with memories at nodes, a qubit of an \EP 
may wait (in a quantum memory) for its counterpart to become 
available so that an \ES operation can be performed. 
In such approaches, as mentioned in the previous section, 
the goal is to determine a swapping tree that 
essentially represents the {\em order} in which \ES operations must be performed
along the selected entanglement path.
In particular,~\cite{Caleffi} formulated the entanglement generation rate 
on a given path between two nodes and then used the formulated path metric
to select the optimal path exhaustively; their path metric is based on 
balanced swapping trees.
In more recent work,~\cite{DPpaper} develops a dynamic programming approach to
determine the swapping tree with minimum expected generation latency, under node and decoherence constraints, for generating an \EP over a given source-destination pair.

\begin{figure}
    \centering
    \includegraphics[width=0.45\textwidth]{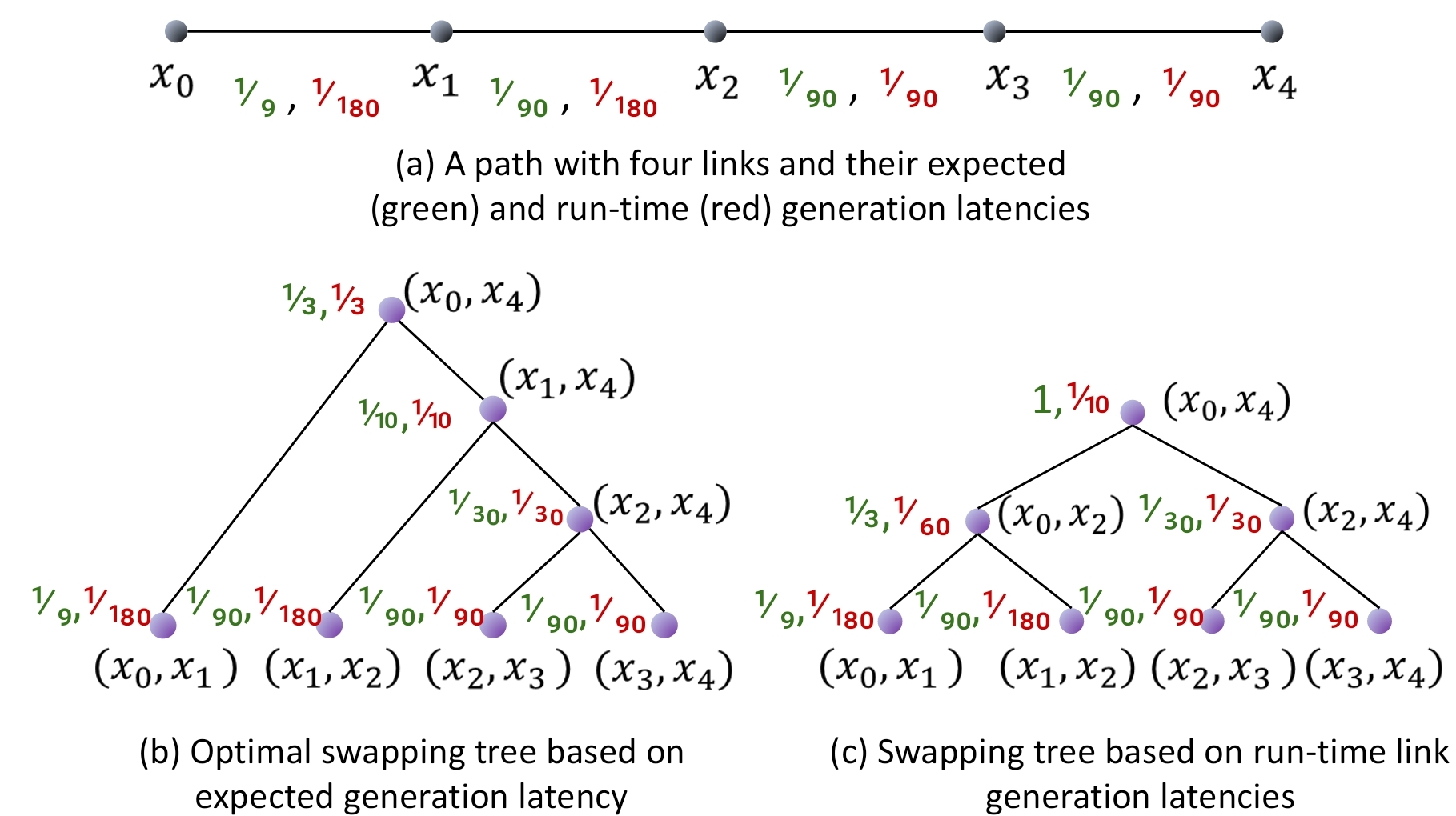}
    \caption{(a). Path $(x_1, \ldots, x_4)$ with expected (green) and actual (at runtime, in red) generation latencies of the link-\EPs. (b) Optimal swapping tree that minimizes the expected generation latency of the \EP over $(x_1, x_4)$. Here, the green values are the expected generation latencies of the \EP at each vertex, and the red values are the actual generation latencies. (c) Alternate swapping tree with sub-optimal {\em expected} generation latency but with lower runtime generation latency compared to the swapping tree in (b).}
    \label{fig:StaticvDynamic}
    \vspace{-0.5cm}
\end{figure}

\softpara{Real-Time Determination of Swapping Order.}
The above works seek to minimize the expected latency of generating
the target \EP by selecting one or more {\em static} swapping trees up front.
However, as discussed in the previous section, an optimal entanglement route
or swapping tree may incur high generation latency at runtime 
depending on the outcome of the link-\EP generation attempts and/or 
\ES operations. Thus, an adaptive strategy that determines the (order of) \ES operations in runtime can outperform fixed swapping trees (see 
Example~\ref{fig:StaticvDynamic}).
With the above insight, a recent work~\cite{MDP} considers 
the problem of minimizing the runtime latency of generating an \EP 
over a pair of nodes using a {\em given} path.
They formulate the problem as a Markov Decision Process 
wherein each state corresponds to each possible network state 
(available \EPs with ages), and develop an optimal 
state transition policy. 
However, their algorithm takes exponential time and space---and 
is feasible only for very small networks (at most 5-10 nodes) 
and low decoherence times (see \S\ref{sec:eval}). 
Our work is inspired by the above work.


\section{\bf \LS Algorithm}
\label{sec:ls_algo}

We describe our proposed \LS algorithm for the \spsw problem. We start by describing a high-level idea of the algorithm. 

\para{Basic Idea.} 
The basic idea behind \LS is to first select a path of minimum expected generation latency over the  $(s, d)$ pair, and, then, at each event (as defined below), pick the \ES $X$ that minimizes the expected latency (after performing $X$) of generating an \EP across $(s, d)$.  
More elaborately, we first use the optimal dynamic programming algorithm of~\cite{DPpaper} to determine the swapping tree with the lowest expected latency of establishing an \EP across $(s,d).$ 
We select the path $P$ corresponding to the leaf nodes of this optimal swapping tree, 
as the entanglement path to generate the \EP over $(s,d)$.
Initially, we start generating link-\EPs over network links on $P$. 
Then, whenever a (link or subpath) \EP over this path $P$ is generated or destroyed
or a sufficient amount of time has passed, we make a determination to either 
perform an \ES operation or {\em wait}.
This determination is made by considering all possible \ES operations available at the given point of time, and for each potential \ES $X$, estimating the minimum latency of generating the target \EP over $(s,d)$ if the \ES $X$ is performed. 
Before presenting the pseudo-code, we discuss two concepts.



\para{Active Links and Trigger Events.}
A link may be \emph{active} or {\em inactive}; an active link continuously attempts to generate a link-\EP over it. Our motivation for inactivating links, as needed, is two-fold: (i) Inactivating a link limits the memory requirements at each network node to at most two; (ii) Allowing a link to continue to generate link-\EPs while a derived \EP exists---leads to existence of multiple \EPs over a pair of nodes with potentially different ages; handling such a generalized network state in our algorithm above is very challenging, and is deferred to our future work.

In our context, the \emph{trigger events} are: (i) Generation of a new \EP, (ii) 
Destruction of an \EP (due to decoherence or \ES failure), 
(iii) {\em Passage of ``sufficient'' time} after a trigger event. 
The motivation behind the last trigger event is to re-determine, based on the updated ages of available \EPs, if some pair of available \EPs should be swapped.  
We thus pick the ``sufficient time'' to be $50 \mu s$, which is equal to half of the time it takes to {\em attempt} a link-\EP generation. Thus, if $50 \mu s$ passes without an event being triggered, then we trigger the {\em passage of sufficient time} event.

\para{\LS Algorithm Pseudo Code.}

\softpara{Inputs:} (i) A quantum network, a pair of network nodes $(s,d)$.

\softpara{Output:} An \EP over $(s,d)$ with minimum generation latency.

\softpara{Algorithm.}
\begin{enumerate}
    \item Determine the routing path $P$ from the optimal swapping tree over $(s,d)$ obtained using \simDP from~\cite{DPpaper} (see \S\ref{sec:gendp}).
    \item Turn all links on $P$ active. 
    \item Continuously (try to) generate a link-\EP across all active links in $P$.
    \begin{itemize}
        \item Once a link-\EP $E$ is generated over a link $e$,  make $e$ inactive. The link $e$ is activated again when either $E$ or an \EP derived from $E$ (via \ESs) is destroyed (due to decoherence or \ES failure). 
    \end{itemize}
    \item On one of the trigger events (as defined above), do:
    \begin{enumerate}
        \item Call \midDP (see below) to determine which of the available pair of \EPs to swap or {\em wait} (for the next trigger event).
        \item Perform the \ES operation or {\em wait} based on (a).
    \end{enumerate}
\end{enumerate}


\para{\bf \midDP Subroutine}.

\softpara{Inputs:} (i) Set of available \EPs (with ages of their qubits), and (ii) Set of active links and the duration for which they have been generating \EPs.

\softpara{Outputs:} Either a pair of available \EPs that should be \ES-ed next, or {\tt wait}.

\softpara{Algorithm.}
\begin{enumerate}
    \item Let $\mathcal{E}\leftarrow$ \{Set of available \EPs\}\ $\cup$\ \{\EPs corresponding to active links\}  
    \item For each pair of \EPs $(E_1, E_2)$ from $\mathcal{E}$ that can be swapped, do:
    \begin{enumerate}
        \item Using \genDP (\S\ref{sec:gendp}), determine the minimum expected latency $L_{E_1,E_2}$ of generating the final \EP over $(s,d)$, if $(E_1,E_2)$ \EPs are swapped. 
    \end{enumerate}
    \item Pick $(E_1,E_2)$ corresponding to the lowest $L_{E_1,E_2}$. If both $E_1$ and $E_2$ are available, return $(E_1,E_2)$, else return {\tt wait}.
\end{enumerate}

\subsection{\bf \genDP: Estimating Minimum Latency}
\label{sec:gendp}

In~\cite{DPpaper}, the authors develop a dynamic programming (DP)
algorithm to determine an optimal (i..e, with minimum expected 
generation
latency) swapping tree to generate an \EP over a given source-destination pair in a quantum network. For our purposes of 
estimating the minimum latency, given a set of available \EPs
and the next \ES operation, we need to modify the DP algorithm
of~\cite{DPpaper} appropriately. 
We briefly describe the original DP algorithm of~\cite{DPpaper} 
and then discuss its modification. 

\para{\DP Algorithm~\cite{DPpaper}.} 
Consider a quantum network with nodes $V=\{1, 2, \ldots, n\}$, 
and a pair $(s,d)$ of source-destination nodes over 
which we want to generate an \EP with minimum expected latency. 
The work in~\cite{DPpaper} develops a dynamic programming (DP)
algorithm (referred to
as \DP, here) that constructs a swapping tree $X$ over a path 
connecting
$s$ and $d$ such that $X$ incurs minimum generation latency.
The \DP algorithm is based on a recursive formula governing the
generation latency of swapping trees under certain 
reasonable assumptions.
In particular, let $T[i,j,h]$ represent the optimal expected latency of 
generating \EP pairs over $(i,j)$ using a swapping tree of height at most $h$.  
Then, we have: 
\begin{align*}
T[i,j,0] &= \frac{\gt}{\gp^2\php} \text{ for adjacent nodes } (i,j).\\
T[i,j,h]&=\min (T[i,j,h-1], (\frac{3}{2}B + \bt)/\bp), \text{ where} \\
B&=\min \limits_{k \in V} \max (T[i,k,h-1],T[k,j,h-1])
\end{align*}
The above equations can be used to compute the optimal generation latency as
well as the corresponding optimal swapping tree. 
The authors in~\cite{DPpaper} also incorporate decoherence and other constraints
in the above algorithm. In particular, to incorporate decoherence constraints, \cite{DPpaper} introduces four additional parameters to the above function $T[]$,
corresponding to the depths of the four grandchildren of a swapping tree's root. 
We refer the reader to~\cite{DPpaper} for more details.


\para{\genDP: Incorporating Ages of Available EPs.} 
The above-described DP starts with a ``base state'' wherein the network links have not yet started generating the corresponding link-\EPs. In contrast, in our context, we need
to estimate the minimum generation latency of the target \EP (see Line~2(a) of Algorithm \midDP) when some of the \EPs are already available---and more 
importantly, may be of non-zero ages. 
To estimate the optimal generation latency for such a generic network state (i.e., with already available \EPs), we divide the entire possible age-range $[0, \tau]$ of a qubit into sufficiently many (we use 100) subranges and estimate the minimum generation latency of the target \EP for each age-range recursively (using dynamic programming) as described  below.
In particular, let $T[i,j,h, a]$ represent the minimum expected latency of 
generating an \EP pair over $(i,j)$ using a swapping tree of height at most $h$ such that the age of the target \EP is within the age-interval $a$. Then, we have: 
\begin{align*}
&T[i,j,0,a] = 0, \text{if}\ (i,j)\ \text{is an available \EP of age in interval $a$}.\\
&T[i,j,0, [0,x]] = \frac{\gt}{\gp^2\php} - t \text{ for adjacent nodes  (i,j)\  \text{where $t$}}\\
& \hspace{0.5in} \text{ is the time already spent generating a link-\EP }\\
& \hspace{0.5in} \text{over $(i,j)$ and  $[0,x]$ is the first subrange of $[0,\tau]$.}\\
&T[i,j,h,a]=\min (T[i,j,h-1,a], (\frac{3}{2}B + \bt)/\bp), \text{ where} \\
&B=\min \limits _{k \in V, a_i \leq a, a_j \leq a} \max (T[i,k,h-1,a_i],T[k,j,h-1,a_j])
\end{align*}

\section{\bf Multi-Path Generalization}
\label{sec:MultiPath}
In this section, we consider the generalization of the \spsw problem wherein we allow for multiple entanglement paths for generating \EPs over a given $(s,d)$ pair. 

\para{Motivation.} In a large quantum network, there may be multiple paths available for 
independent generation of the desired remote \EP, and the stochasticity of the processes 
imply
that using multiple paths can result in lower expected latency (any one of the paths 
needs to succeed). 
Thus, many works have considered multi-paths~\cite{DPpaper, ghaderibaneh2023generation} for entanglement generation. 
In our context, we generalize the \spsw problem for multi-paths, i.e.,
for real-time determination of entanglement path and \ES operations during 
the course of \EP generation. 

\para{\mpsw Problem (Multi-Path Swapping Order).} 
Given a quantum network and a source-destination pair $\{(s, d)\}$, 
the \mpsw problem is to determine the choice of entanglement path and ordering of \ES operations (in real-time as the \EPs are generated) so as to minimize the actual/real-time generation latency of an \EP $(s,d)$ under the decoherence and node constraints (as defined in \spsw problem formulation).  


\para{Multi-Path Optimal Static Approach.} 
In recent work,~\cite{DPpaper} presents a linear programming (LP) based algorithm (referred to as \LP, here) for constructing an ``aggregated'' structure of multiple swapping trees that yields an optimal generation rate of \EP over a given source-destination pair. Their technique is based on constructing an appropriate hypergraph
that has embedded in it all possible swapping trees in the given quantum network, and 
then developing a linear program to determine the optimal hyperflow in the constructed hypergraph; the optimal hyperflow essentially represents the optimal ``aggregated'' tree
structure for the generation of \EPs over the given source-destination pair. 
We use the above approach as a baseline and develop a real-time \ES-ordering 
approach as described below.
    
\para{\LSMP Algorithm for \mpsw.} We utilize the \LP algorithm above first to obtain multiple paths (not necessarily disjoint) 
from $s$ to $d$ that can/should be used to establish \EPs $(s,d)$; these paths
can be obtained by extracting a sufficient number of swapping trees from the \LP ``aggregated tree'' solution. 
Then, we proceed similarly to the previous \LS algorithm for the single path.
In particular, as in \LS, we activate all the links and continuously
 attempt to generate link-\EPs over them. 
Then, on any trigger event (as defined before), we call the \midDP procedure
on each of the paths {\em independently} to determine the best pair of \EPs to swap (or wait) on each path. Since the paths may not be disjoint, the best pair of \EPs chosen for a path may involve \EPs that are part of multiple paths.
Thus, we sort these best pairs from the paths based on the expected generation latencies of the corresponding paths and perform the swaps in that order (as long as the required EPs haven't been consumed by a previous swap). We describe this
formally in the below pseudo code. The algorithm terminates when an \EP has been established across $(s,d)$. 


\para{\LSMP Psuedo-Code.}

\softpara{Inputs:} (i) A quantum network, and a source-destination pair $(s,d)$. 

\softpara{Output:} An \EP over the nodes $(s,d)$ in minimum latency.

\softpara{Algorithm.}
\begin{enumerate}
    \item Select a set of routing paths $\{P_i\}_{[k]}$ using \LP, as described above. 
    \item Turn on all links in each $P_i$ as active.
    \item Continuously (try to) generate a link-\EP across all active links in each $P_i.$
    \begin{itemize}
        \item Once a link \EP $E$ is generated over a link $e$,  make $e$  inactive. The link $e$ is activated again when either $E$ or an \EP derived from it (via \ESs) is destroyed (due to decoherence or \ES failure). 
    \end{itemize}
    \item On one of the trigger events, do
    \begin{enumerate}
        \item For each path $P_i$ independently, use \midDP (\S\ref{sec:ls_algo}) to determine the pair of \EPs $D_i$ to be swapped next. Let $L_i$ be the expected generation latency of the target \EP over path $P_i$ after performing $D_i$. If $D_i$ is null (i.e., when {\tt wait} is returned), then let $L_i$ be infinity.
        \item Traverse the $D_i$'s in ascending order of $L_i$'s, and perform the \ES corresponding to each $D_i$ only if neither of the \EPs in $D_i$ have been consumed by a $D_j$. 
    \end{enumerate}
\end{enumerate}

\section{\bf Evaluation}
\label{sec:eval}

This section evaluates our algorithm on different network graphs with varying parameters.

\para{Algorithms Compared.} For the \spsw problem, we compare four algorithms: (i) \LS from~\S\ref{sec:ls_algo}; (ii) Optimal dynamic programming algorithm (\simDP) from~\cite{DPpaper} which returns an optimal static swapping tree; (iii) \Asap from~\cite{MDP}, which performs an \ES operation over EPs as soon as any are available; (iv) Markov Decision Process based (\MDP) algorithm from~\cite{MDP}. 
For the multi-path \mpsw problem, we compare two algorithms: (v) \LSMP from~\S\ref{sec:MultiPath}; (vi) \LP from \cite{DPpaper} (see \S\ref{sec:MultiPath}).

\begin{figure}[t]
\begin{subfigure}{\linewidth}
    \centering
    \includegraphics[width=\textwidth]{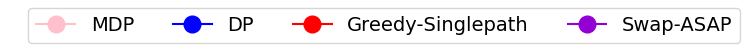}
\end{subfigure}\\
\begin{subfigure}{0.49\linewidth}
\centering
\includegraphics[width=\textwidth]{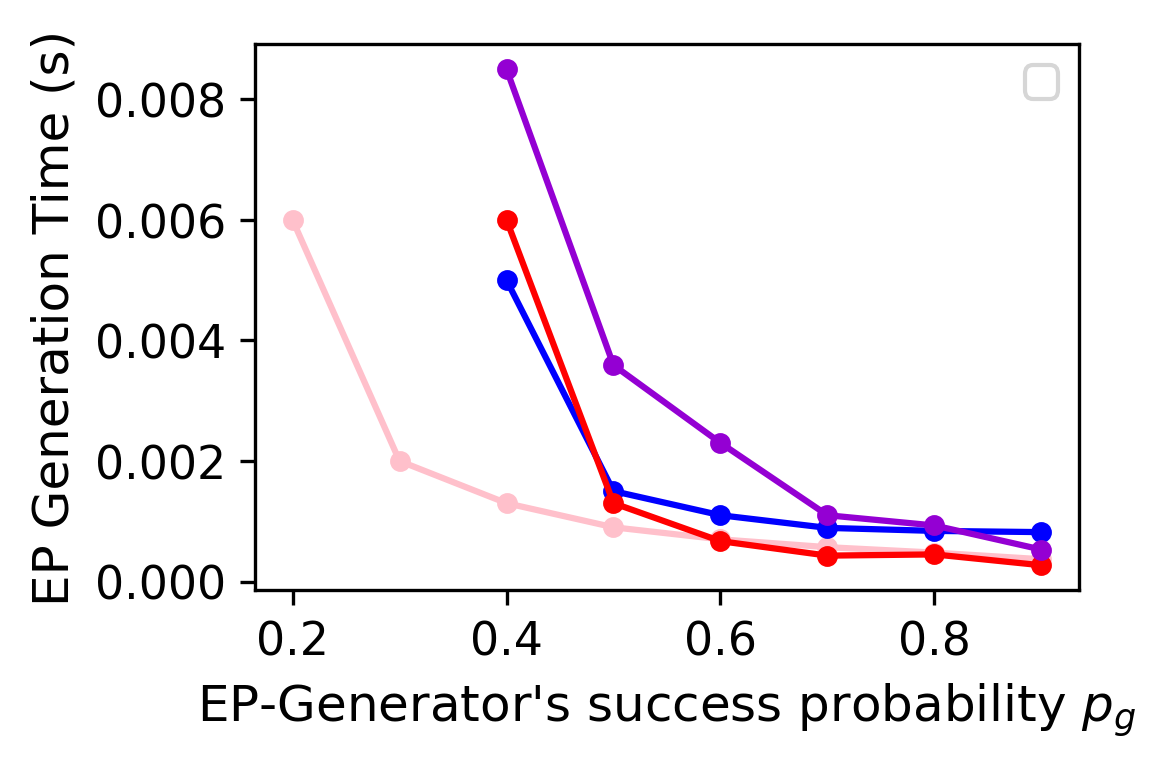}
\vspace{-0.6cm}
\captionlistentry{}
\label{fig:cutoff5}
\end{subfigure}
\begin{subfigure}{0.49\linewidth}
\centering
\includegraphics[width=\textwidth]{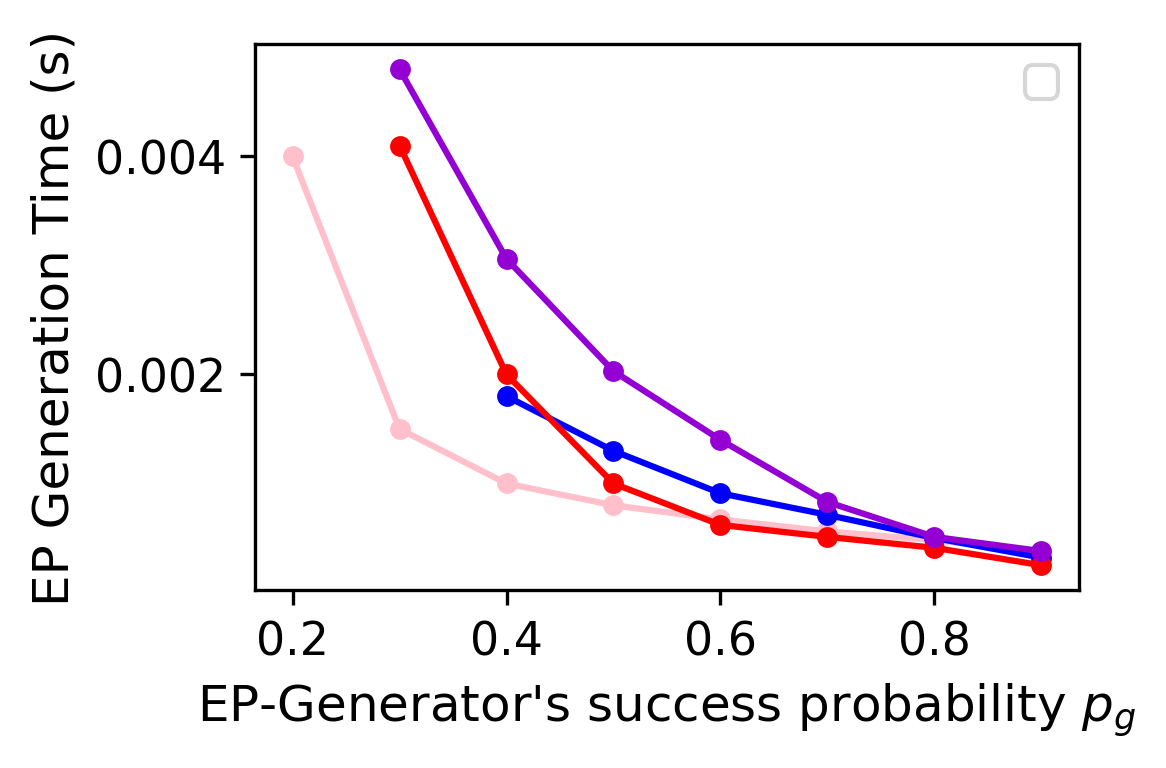}
\vspace{-0.6cm}
\captionlistentry{}
\label{fig:cutoff10}
\end{subfigure}
\caption{\spsw (single-path) Problem. \EP generation latency of various algorithms, for small networks (5 nodes) and small decoherence times (\num{1.5e-4} (left) and \num{3e-4} (right) seconds), for varying node EP-generator's success probabilities.}
\vspace{-0.5cm}
\end{figure}

\begin{figure*}[!h]
\vspace{-0.2in}
\begin{subfigure}{\linewidth}
    \centering
    \includegraphics[width=0.5\textwidth]{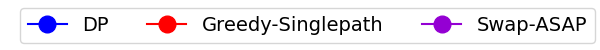}
\end{subfigure}\\
\begin{subfigure}{0.33\linewidth}
\centering
\includegraphics[width=\textwidth]{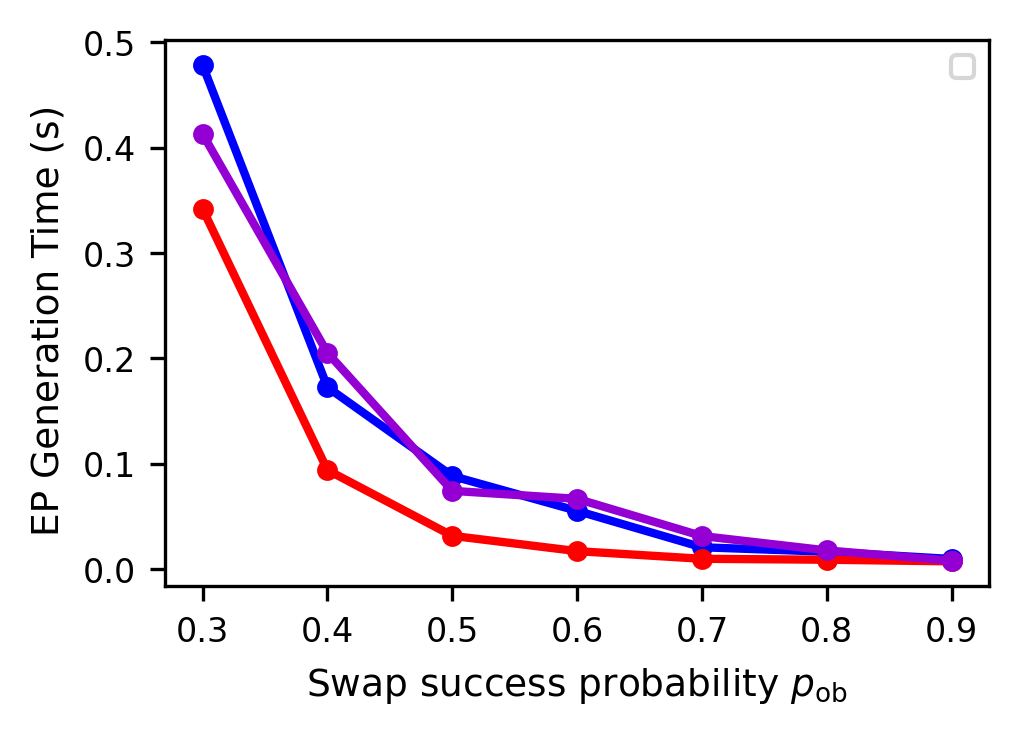}
\vspace{-0.6cm}
\captionlistentry{}
\label{fig:VarySwap}
\end{subfigure}
\hspace{-0.4cm}
\begin{subfigure}{0.33\linewidth}
\centering
\includegraphics[width=\textwidth]{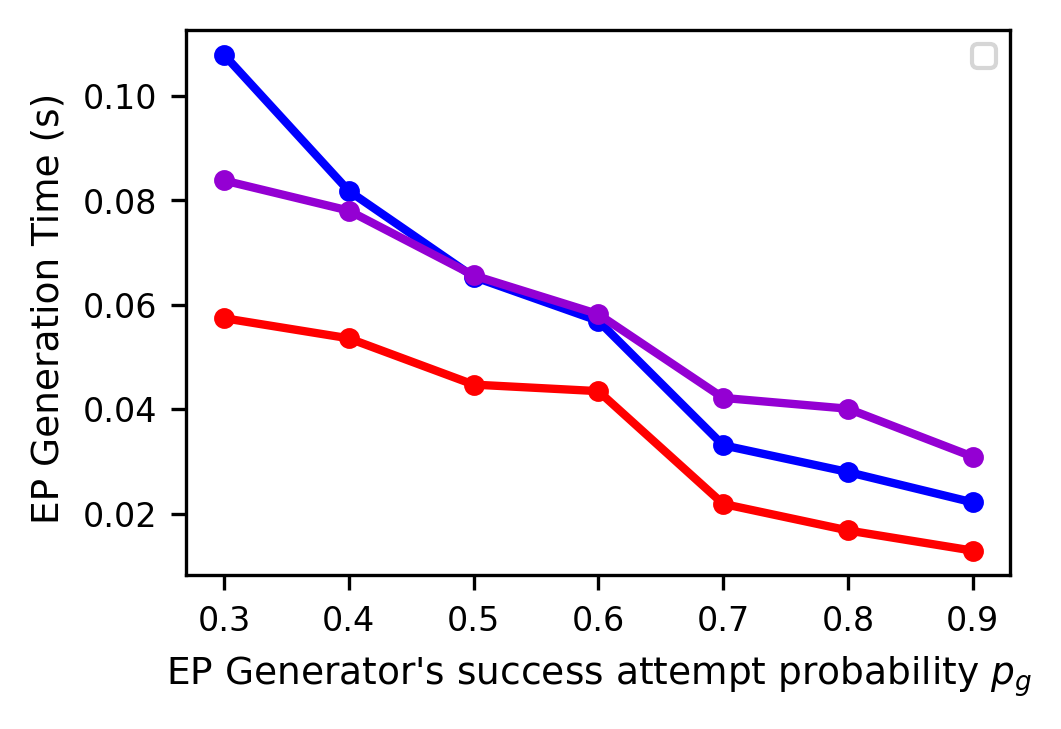}
\vspace{-0.6cm}
\captionlistentry{}
\label{fig:VaryNodeProb}
\end{subfigure}
\hspace{-0.4cm}
\begin{subfigure}{0.33\linewidth}
\centering
\vspace{-0.2cm}
\includegraphics[width=\textwidth]{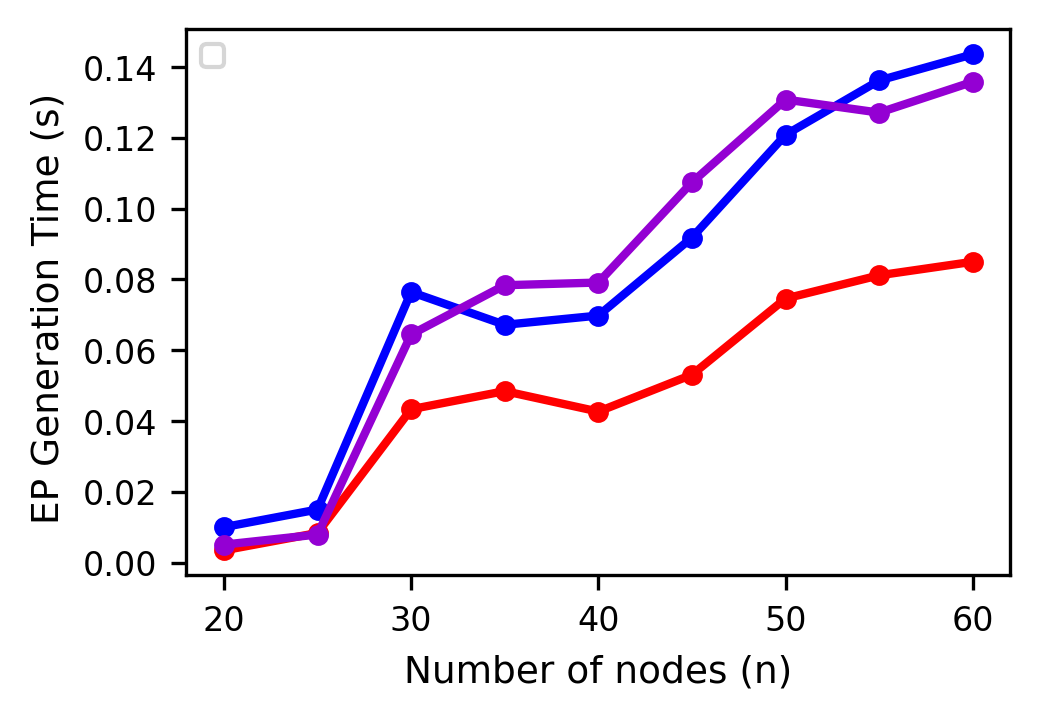}
\vspace{-0.6cm}
\captionlistentry{}
\label{fig:VaryNumNodes}
\end{subfigure}
\begin{subfigure}{0.33\linewidth}
\centering
\includegraphics[width=\textwidth]{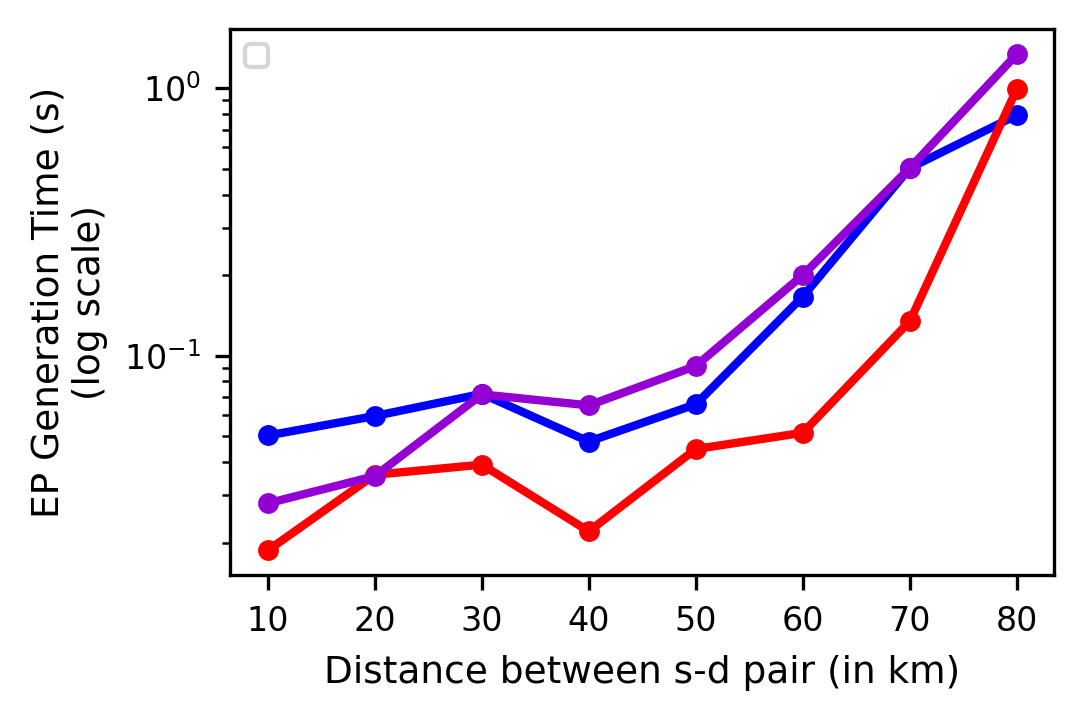}
\vspace{-0.6cm}
\captionlistentry{}
\label{fig:Varydist}
\end{subfigure}
\hspace{-0.3cm}
\begin{subfigure}{0.63\linewidth}
\centering
\includegraphics[width=\textwidth]{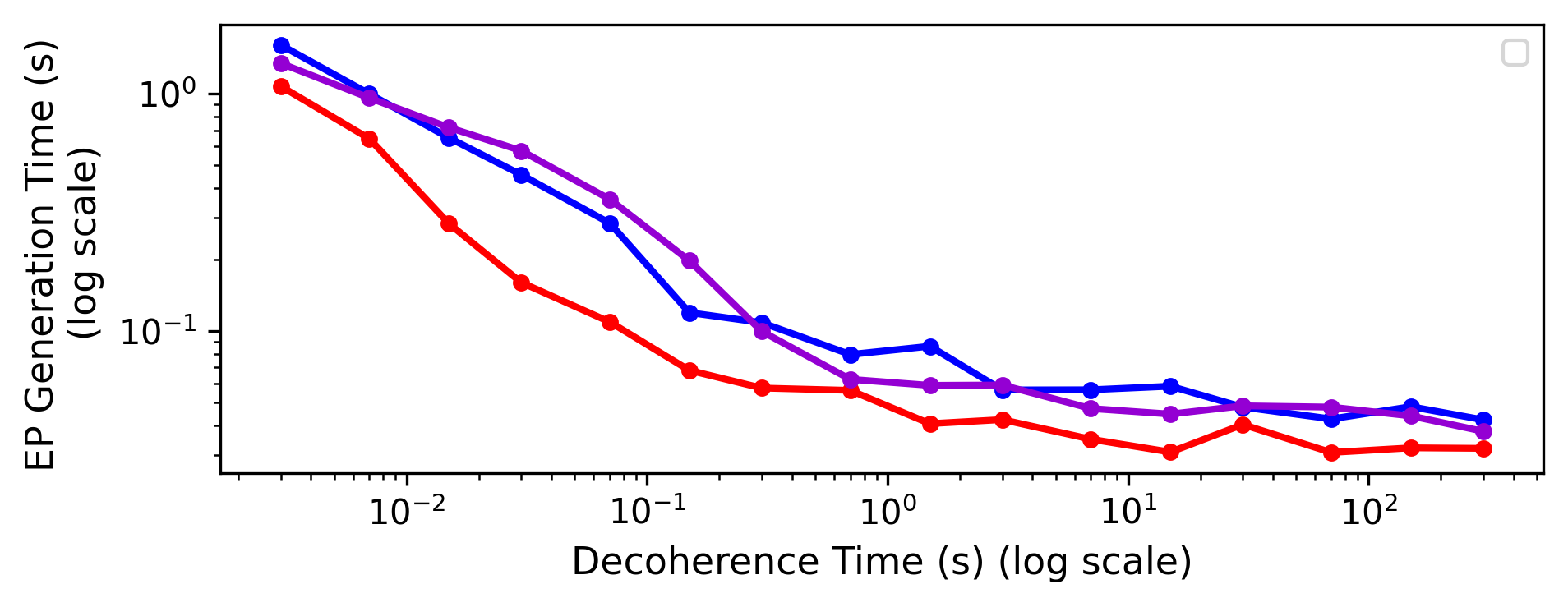}
\vspace{-0.8cm}
\captionlistentry{}
\label{fig:VaryCutoff}
\end{subfigure}
\caption{\spsw (single-path) Problem. \EP generation latency of various algorithms, \Asap, \DP, and \LS, for varying parameters. \MDP is not shown due to its prohibitive runtime for these parameter values (see text).}
\vspace{-0.3cm}
\end{figure*}

\begin{figure*}[h]
\begin{subfigure}{0.35\linewidth}
\centering
\includegraphics[width=\textwidth]{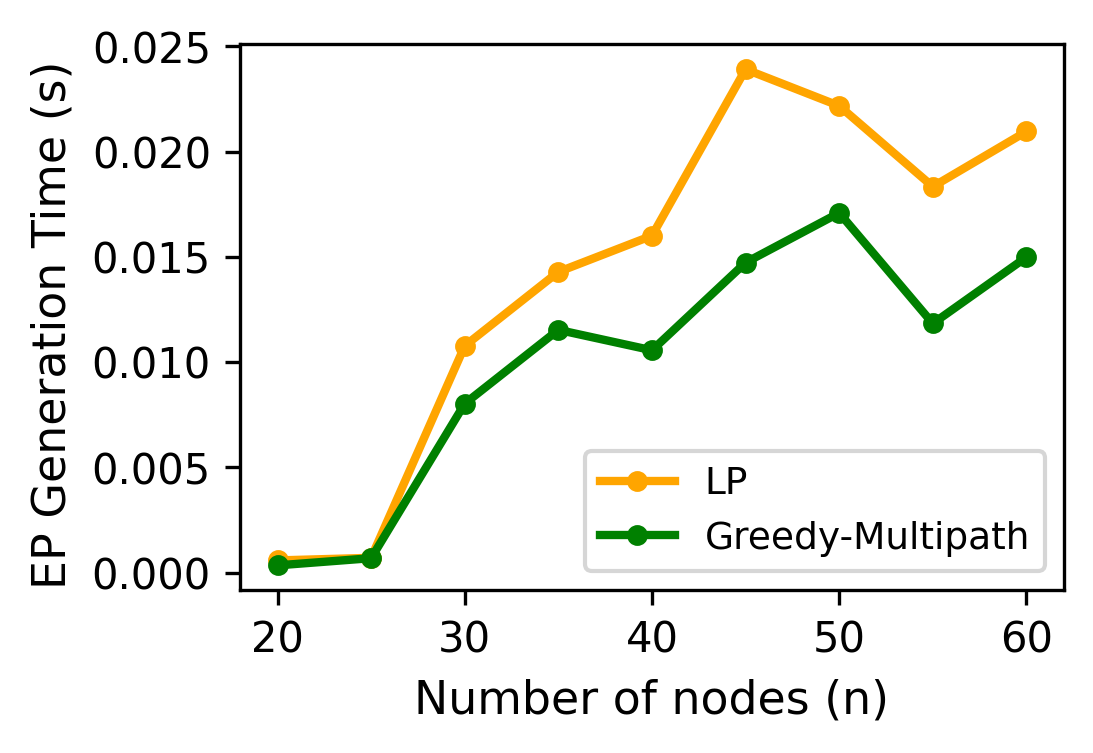}
\captionlistentry{}
\label{fig:GreedyvsLPNumNodes}
\end{subfigure}
\begin{subfigure}{0.63\linewidth}
\centering
\includegraphics[width=\textwidth]{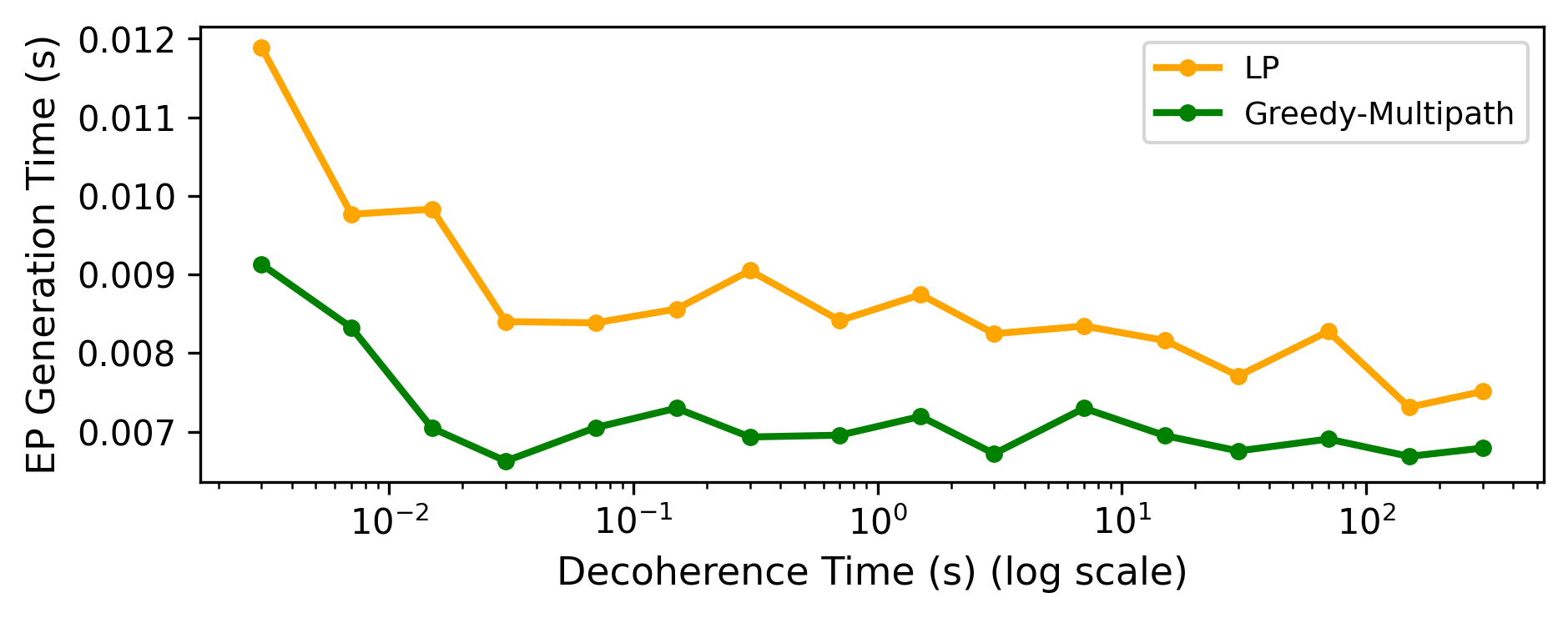}
\captionlistentry{}
\label{fig:GreedyvsLPcutoff}
\vspace{-0.1cm}
\end{subfigure}
\vspace{-0.2in}
\caption{\mpsw (multi-path) Problem. \EP generation latencies of \LSMP and \LP for varying parameters. }
\vspace{-0.2in}
\end{figure*}

\para{Implementation Details.} 
Given a quantum network and a single $(s,d)$ pair, \simDP picks a {\em static} swapping tree of minimum expected latency and performs swap operations in 
the order given by the tree. 
\MDP maintains a state of all possible \EPs over all possible ages, resulting in a state space of size $\mathcal{O}(t^n)$, where $n$ is the number of nodes in the network and $t=2\tau/3\gt$ is a discretized representation of the decoherence threshold. It then pre-computes a policy that, for each state, returns an action: 
{\em wait} or perform a certain swapping operation. 
Given a network path and an objective of establishing an \EP across the end nodes, \Asap continuously attempts link generation of active links. When a pair of  \EPs that can be swapped become available, \Asap swaps them. 
Note that both \MDP and \Asap take a network path rather than a network graph as input; thus, as for the case of \LS, they operate on the path corresponding to the optimal swapping tree of \simDP.
In all approaches, the links become active or inactive as in our approaches \LS and 
\LSMP.



\para{Generating Random Networks.} We use a quantum network spread over an area of $100km \times 100km$. We use the Waxman model~\cite{waxman1988routing}, which has been used to create Internet topologies. We vary the number of nodes in the network from $20$ to $60$, with $40$ as the default value. 
We generate random networks by varying the following parameters.
\begin{itemize}
    \item Number of nodes (default = $40$)
    \item Swapping success probability \php (default = 0.5)
    \item Node \EP-generator's success probability \gp (default = 0.5)
    \item Decoherence threshold $\tau$ (default = $1.5 s$)
    \item Physical distance between the source-destination pair $s$ and $d$ (default = 20-50km).
\end{itemize}
In all graphs, each data point is the average generation latency of $10$ runs of the corresponding algorithm on a set of input parameters. 

\para{Evaluation Results.} For the single-path (\spsw) problem, we evaluate the four algorithms on inputs generated by varying the parameters described above. We note that \MDP takes prohibitively long to run on paths with more than $7$ nodes or settings with
decoherence time more than \num{3e-4} sec with paths of more than $5$ nodes.
Thus, we evaluate \MDP on small networks with $5$ nodes and two decoherence times, viz., 0.15ms and 0.3ms.
See Figs~\ref{fig:cutoff5}-\ref{fig:cutoff10}. For such \underline{small networks}, we observe that \MDP performs the best, as expected, as it is the optimal algorithm. However, we note that \LS also performs very close to the optimal \MDP for $\gp \geq 0.6$ values.
For \underline{larger networks} (Figs~\ref{fig:VarySwap}-\ref{fig:VaryCutoff}), we observe that \LS consistently performs better than \simDP and \Asap for most instances; in particular, it achieves generation latencies of up to $40\%$ lower 
than the optimal static approach (\simDP) and up to $45\%$ lower than the naive adaptive approach \Asap.
Finally, for the \underline{multi-path} (\mpsw) problem, we compare the adaptive \LSMP and the static-optimal \LP approaches. See Figs.~\ref{fig:GreedyvsLPNumNodes}-\ref{fig:GreedyvsLPcutoff}. We observe that \LSMP performs 10-20\% better 
than \LP in most cases. 

\para{Runtime Overhead.} The running time of the selection phase of our
\LS and \LSMP algorithms is 100 $\mu$s and 700 $\mu$s for the
default parameter values on an 8-core Intel Core i7-11700 system. Since
the selection phase involves computing the expected latency for each 
possible choice independently---the running time can be easily 
parallelized and reduced to the order of a few $\mu$s.
Thus, the running overhead of our approaches has minimal impact on 
the decoherence of qubits (since realistic values of decoherence times 
is of the order of several seconds) and generation latency (which is
of the order of 10s-100s of milliseconds).

\section{\bf Conclusion}

In this work, we have developed an adaptive EP generation algorithm that chooses
the entanglement route and/or swapping operations at EP-generation time, based on 
the network state at the time. In our future work, we seek to extend our work in the following ways: (i) Consider multiple source-destination pairs; (ii) Continuous generation of EPs with node-memory constraints; (iii) Consider discarding EPs
that are likely to decohere before the target \EP can be generated.



\bibliography{ref}
\end{document}